\documentclass{sig-alternate}

\usepackage{cite}     
\usepackage{epsfig}     
\usepackage{url}        
\usepackage{subfigure} 
\usepackage{version} 

\begin{document}
%
\title{On TCP-based SIP Server Overload Control}



\author{Charles Shen and Henning Schulzrinne \\
Department of Computer Science, Columbia University \\
New York, NY 10027 \\
\{charles,hgs\}@cs.columbia.edu}

\conferenceinfo{IPTComm 2010, 2-3 August, 2010}{Munich, Germany} \CopyrightYear{2010} \crdata{}


%




\includeversion{cf-only}
\excludeversion{tp-only}

\maketitle


%
%
%
%

\begin{abstract}

\begin{cf-only}
The Session Initiation Protocol (SIP) server overload management has
attracted interest since SIP is being widely deployed in the Next Generation
Networks (NGN) as a core signaling protocol. Yet all existing
SIP overload control work is focused on SIP-over-UDP, despite the fact that TCP
is increasingly seen as the more viable choice of SIP transport. This
paper answers the following questions: is the existing
TCP flow control capable of handling the SIP overload 
problem? If not, why and how can we make it work? We provide a
comprehensive explanation of the default SIP-over-TCP overload
behavior through server instrumentation. We also propose and implement
novel but simple overload control algorithms without any kernel or
protocol level modification. Experimental evaluation shows that with our
mechanism the overload performance improves from its original zero
throughput to nearly full
capacity. Our work leads to the important general insight
that the traditional notion of TCP flow control alone is incapable of
managing overload for time-critical session-based
applications, which would be applicable not only to SIP, but also to a wide range of other common applications such as database servers.   
\end{cf-only}

\end{abstract}

\section{Introduction}
\label{sec-intro}

The Session Initiation Protocol (SIP)~\cite{rfc3261} is an application layer signaling protocol for 
creating, modifying, and terminating media sessions in the Internet.  
SIP has been adopted by major standardization bodies including 3GPP,
ITU-T, and ETSI as the core signaling protocol of Next
Generation Networks (NGN) for services such as Voice over IP (VoIP), conferencing, 
Video on Demand (VoD), presence, and Instant Messaging (IM). The
increasingly wide deployment of SIP has raised a requirement for 
SIP server overload management solutions~\cite{rfc5390}. SIP server can be overloaded for many reasons such as emergency-induced call volume, flash crowds generated by TV programs
(e.g., American Idol), special events such as ``free tickets to the third
caller'', or denial of service attacks. 

\begin{cf-only}
Although a SIP server is an application server, the SIP
server overload problem is distinct from other well-known application server
such as HTTP overload because in the SIP architecture, multiple server
hops are common. There are also many SIP application level retransmission timers, and there is a time-critical session completion requirement. SIP's built-in session
rejection mechanism is known to be unable to manage
overload~\cite{rfc5390} because it could cause the server to spend
all cycles rejecting messages and result in congestion collapse.  
If, as often recommended, the rejected sessions are sent to a load-sharing SIP
server, the alternative server will soon also be generating nothing
but rejection messages, leading to a cascading failure.
Hilt {\em et al.}~\cite{hilt:overload,hilt:overload-design} articulate a SIP
overload control framework based on augmenting the current SIP
specification with application level feedback between  
SIP proxy servers. The feedback,
which may be rate-based or window-based, pushes the burden
of rejecting excessive sessions from the target server
to its upstream servers and thus prevents the overload. Detailed SIP application
level feedback algorithms and their effectiveness have been
demonstrated by a number of researchers, e.g., Noel~\cite{noel07}, Shen~\cite{shen08} and Hilt~\cite{hilt08}. 

As far as we know, all existing SIP overload control design and evaluation
focus on SIP-over-UDP, presumably because UDP is still the common
choice for today's SIP operational environment. However, SIP-over-TCP
is getting increasingly popular 
and seen as a more viable SIP transport choice for a number of
reasons,
such as the need for securing SIP
signaling over
TLS/TCP~\cite{voipsecurity,shen09techreport,sipforum,rfc3261}
(There is also a newer TLS version - Datagram TLS, which runs over
UDP, but its deployment popularity is not clear), 
support for message sizes exceeding the maximum UDP datagram
size~\cite{rfc3261}, facilitation of firewall and NATs traversal~\cite{kumiko08}, and potentially overload control. 

\end{cf-only}

The SIP-over-TCP overload control problem differs in two main  
aspects from the SIP-over-UDP overload control
problem. One is TCP's built-in flow control
mechanism which provides an inherent, existing channel for
feedback-based overload control. The other is the
removal of many application layer retransmission timers that exacerbates the overload
condition in SIP-over-UDP. 
Nahum {\em et al.}~\cite{erich07} have experimentally studied SIP
performance and found that overload leads to congestion collapse
 for both SIP-over-TCP and SIP-over-UDP. Their focus, however, is not on overload control so they do not discuss why SIP-over-TCP congestion collapse happens or how to prevent it. Hilt
{\em et al.}~\cite{hilt08} have shown simulation results by applying
application level feedback control to SIP servers with TCP-specific
SIP timers but without including a TCP transport stack in the simulation.

\begin{cf-only}This paper\end{cf-only}
systematically addresses the SIP-over-TCP overload control
problem through an experimental study and analysis. To the authors' knowledge, our paper is the first to provide a
comprehensive answer to the following questions: why is there still
congestion collapse in SIP-over-TCP despite the presence of the well-known TCP
flow control mechanism and much fewer SIP retransmission timers? Is
there a way we can utilize the existing
TCP infrastructure to solve the overload problem without changing
the SIP protocol specification as is needed for the UDP-based
application level feedback mechanisms?   

\begin{cf-only}

We find that the key reasons why TCP flow control feedback does {\em
not} prevent SIP congestion collapse has to do with the session-based
SIP load characteristics and the fact that the session needs to be
established within the timeout threshold. Different messages in the
message flow of the same SIP session arrive at
different times from upstream and downstream SIP entities; start-of-session requests trigger all the remaining in-session
messages and are therefore especially expensive. The transport level connection-based TCP
flow control, without knowing the causal relationship
among the messages, will admit too many start-of-session requests and result
in a continued accumulation of in-progress sessions in the system,
leading to large queuing delays. When that happens, the TCP flow
control creates back pressure propagating to the session originators, adversely
affecting their ability to generate messages that could complete
existing sessions. In the meantime, SIP response retransmission still kicks
in. The combined delayed message
generation and processing as well as response retransmission lead to SIP-over-TCP congestion collapse.

Based on our observations, we propose a novel SIP overload control
mechanisms within the existing TCP flow control infrastructure. To
respect the distinction between start-of-session requests and
other messages, we introduce the concept of {\em connection split}. To meet the delay requirements and prevent
retransmission, we develop {\em smart forwarding} algorithms combined
with {\em buffer
minimization}. Our mechanisms contain only a single tunable parameter
for which we provide a recommended value. Implementation of our
mechanisms exploits existing Linux socket API calls and is extremely
simple. It does not require any modifications at the kernel level,
nor changes to the SIP or TCP specification.

We evaluate throughput, delay and fairness results of our mechanisms on a common Intel-based Linux testbed using
the popular open source OpenSIPS server with up to ten upstream servers  
overloading the target server at over ten times the server capacity. 

Our mechanism is best suited for the common case where the
number of upstream servers overloading the target server at the same time is not excessively
large, such as servers in the core networks of big 
service providers. But we also point out possible 
solutions when a large number of upstream servers overload a
single target server, such as when numerous enterprise servers connect to the
same server from a big service provider.

Our research leads to the important insight that the traditional notion of
TCP flow control alone is insufficient in preventing 
congestion collapse for time-sensitive session-based loads, which
cover a broad range of applications, e.g., from SIP servers to
data center systems~\cite{vasudevan09}.  

\end{cf-only}

The remainder of this paper is structured as follows.
Section~\ref{sec:related} describes related work. 
Section~\ref{sec:background} provides some background on SIP and TCP
flow and congestion control.
Section~\ref{sec:testbed} describes the experimental testbed used
for our experiments.
Section~\ref{sec:results-default} explains the SIP-over-TCP congestion
collapse behavior. 
Section~\ref{sec:ourmec} and Section~\ref{sec:overallperf} develop and evaluate our overload control mechanism.

\section{Related Work}
\label{sec:related}

\begin{cf-only}

SIP overload falls into the broader category of application server
overload where, in particular, web server overload
control~\cite{colajanni98,elnikety04,zhao06} has been studied
extensively. 
Although most of the work on web
server overload control uses a request-based workload model, Cherkasova
and Phaal~\cite{cherkasova02} presented a study using session-based
workload, which is closer to our SIP overload study. However their
mechanism uses the overloaded server to reject excessive loads,
which is known to be insufficient for SIP~\cite{rfc5390}.    
\end{cf-only}

\begin{cf-only}
A number of authors~\cite{erich07,kumiko08,shen09techreport,RamFCR08}
have measured SIP server performance over TCP, without 
discussing overload. 
\end{cf-only}
The SIP server overload problem itself has received intensive attention
only recently. Ejzak {\em et al.}~\cite{ejzak04} provided
a qualitative comparison of the overload in PSTN SS7 signaling
networks and SIP networks. Whitehead~\cite{gocap} described a
protocol-independent overload control framework called GOCAP but its
mapping to SIP is still being defined. 
Ohta~\cite{ohta06} explored the approach of using a
priority queueing and bang-bang type of overload control through
simulation. Noel and Johnson~\cite{noel07} presented initial results
of a rate-based SIP overload control mechanism. Sun {\em et al.}~\cite{sun07} proposed adding a front end SIP flow management
system to conduct overload control including message scheduling,
admission control and retransmission removal. Sengar~\cite{sengar09}
combined the SIP built-in backoff retransmission mechanism with a
selective admittance method to provide server-side pushback for
overload prevention. %
Hilt {\em et al.}~\cite{hilt08} provided a side-by-side comparison
of a number of overload control algorithms for a network of SIP
servers, and also examined different overload control paradigms such
as local, hop-by-hop and end-to-end overload control. Shen {\em et al.}~\cite{shen08} proposed three window-based SIP feedback control
algorithms and compared them with rate-control algorithms. Except
for~\cite{hilt08}, all of the above work on SIP overload control
assumes UDP as the transport. Hilt {\em et al.}~\cite{hilt08} present simulation of application level feedback overload control for SIP server with only TCP-specific SIP timers enabled, but their simulation does not include a TCP transport stack.   

\begin{cf-only}

The basic TCP flow and congestion control mechanisms are
documented in~\cite{rfc793,jacobson88}. Modifications to the basic TCP
algorithm have been proposed to improve various aspects of TCP
performance, such as start-up behavior~\cite{hoe96}, retransmission
fast recovery~\cite{rfc3782}, packet loss recovery
efficiency~\cite{rfc2018,rfc2883}, or overall 
congestion control~\cite{rfc2581,brakmo95}. There are also research efforts
to optimize the TCP algorithm for more recent network architecture
such as mobile and wireless
networks~\cite{elaarag02,wuxiuchao08} and high-speed
networks~\cite{kliazovich08,ha08}, as well as additional work 
that focuses not on modifying TCP flow and congestion control
algorithm itself, but on using dynamic socket buffer tunning methods
to improve
performance~\cite{hasegawa01,dunigan02}.
Another category of related work focuses on routers, e.g., active buffer
management ~\cite{floyd93,morris00} and router buffer
sizing~\cite{vishwanath09}. Our work differs from all the above in
that our metric is not the direct TCP throughput, but the
application level throughput. Our goal is to explore the existing TCP
flow control mechanism for application level overload management,
without introducing TCP or kernel modifications. 

There are also studies on TCP performance for
real-time media, e.g.,~\cite{argyriou07,baset06,wang04}. Our work,
however, addresses the session establishment phase for real-time
services, which has very different load characteristics.

\end{cf-only}

\section{Background}
\label{sec:background}

\subsection{SIP Overview}

\label{sssec:stproxy}

SIP defines two basic types of entities: User Agents (UAs) and servers. 
UAs represent SIP end points. SIP servers can be either registrar servers
for location management, or proxy servers for message forwarding. 
SIP messages are divided into requests (e.g., {\sf INVITE} and 
{\sf BYE} to create and terminate a SIP session, respectively) and 
responses (e.g., {\sf 200 OK} for confirming a session setup). 

\begin{figure}
\centering
\epsfig{file=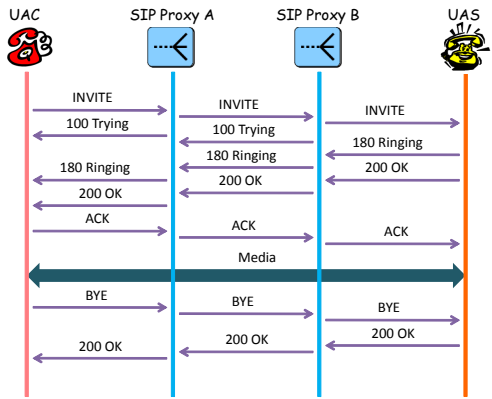, scale=0.4}
\caption{Basic SIP call flow}
\label{fig:sip-call-flow}
\end{figure}

SIP message forwarding, known as proxying, is a critical function of 
the SIP infrastructure.
Fig.~\ref{fig:sip-call-flow} shows a typical message flow of stateful SIP 
proxying where all SIP messages are routed through the proxy with the
SIP \texttt{Record-Route} option enabled. Two SIP UAs, designated as User Agent Client (UAC) and User
Agent Server (UAS), represent the caller and callee of a multimedia
session. The UAC wishes to establish a session with the 
UAS and sends an {\sf INVITE} request to proxy A. Proxy A looks up the contact address for the SIP URI of the UAS 
and, assuming it is available, forwards the message to proxy B, where the UAS can be
reached. Both proxy servers also send {\sf 100 Trying} response to
inform the upstream SIP entities that the message has been 
received. After proxy B forwards the message to the UAS. The UAS
acknowledges receipt of the {\sf INVITE} with 
a {\sf 180 Ringing} response and rings the callee's phone. 
When the callee actually picks up the phone, the UAS sends out a {\sf
200 OK} response. 
Both the {\sf 180 Ringing} and {\sf 200 OK} make their way 
back to the UAC. 
The UAC then generates an {\sf ACK} request for the {\sf 200 OK}.
\begin{cf-only}Having established the session, the media flows directly between the two endpoints.\end{cf-only} When the conversation is finished, the UAC ``hangs up'' and generates 
a {\sf BYE} request that the proxy servers forward to the UAS. The UAS then 
responds with a {\sf 200 OK} response which is forwarded back to the UAC.

\begin{figure} [hbp]
\centering
\epsfig{file=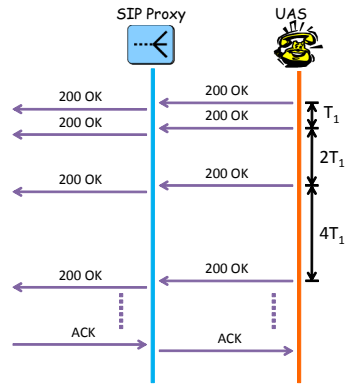, scale=0.4}
\caption{{\sf 200 OK} retransmission}
\label{fig:200retrans}
\end{figure}

SIP is an application level protocol on top of the transport
layer. It can run over any common transport layer protocols, such
as UDP, TCP and SCTP~\cite{rfc4960}. SIP defines quite a number of timers. One group of
timers is for hop-to-hop message retransmissions in case a message is lost. 
These retransmission timers are not used when TCP is the
transport because TCP already provides a reliable transfer. There is 
however a retransmission timer for the end-to-end {\sf 200 OK}
responses which
is enabled even when using TCP transport, in order to accommodate
circumstances where not all links in the path are using reliable
transport.
The {\sf 200 OK} retransmission timer is shown in
Fig.~\ref{fig:200retrans}. The timer
starts with $T_1=500\,ms$ and 
doubles until it reaches $T_2=4\,s$. From then on the timer value
remains at $T_2$ until the total timeout period exceeds 32~s, when the
session is considered to have failed. 
The UAC should generate
an {\sf ACK} upon receiving a {\sf 200 OK}. The UAS cancels the {\sf
200 OK} retransmission timer when it receives a corresponding {\sf ACK}.

\subsection{Types of SIP Server Overload}

There are many causes to SIP overload, but the resulting SIP overload
cases can be grouped into either of the two types:
proxy-to-proxy overload or UA-to-registrar overload.

\begin{figure}[thbp]
  \begin{center}
      \subfigure[proxy-to-proxy overload]{\label{fig:s2soverload}
          \includegraphics[scale=0.5]{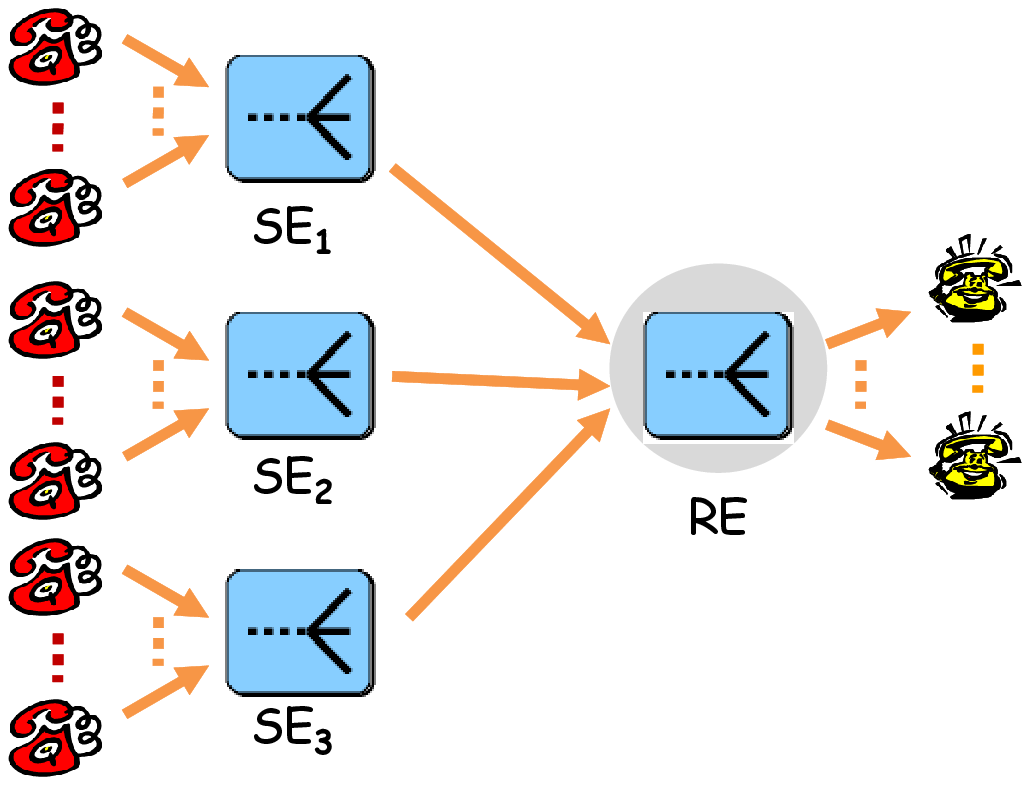}}
      \subfigure[UA-to-registrar overload]{\label{fig:c2soverload}
        \includegraphics[scale=0.4]{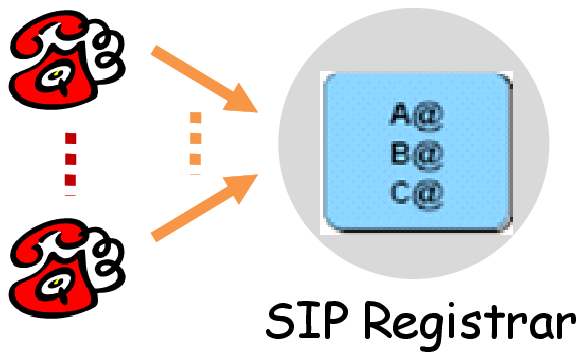}}
    \caption{Types of SIP server overload}
    \label{fig:-screenlog}
  \end{center}
\end{figure}

A typical proxy-to-proxy overload topology is illustrated in
Fig.~\ref{fig:s2soverload}, where the
overloaded proxy server is connected to a
relatively small number of upstream proxy servers. The overloaded
server in Fig.~\ref{fig:s2soverload} is also referred to as a Receiving
Entity (RE) and its upstream servers are also referred to as Sending
Entities (SEs)~\cite{hilt:overload-design}. One example of the proxy-to-proxy overload is a special event
like ``free tickets to the third caller'', also known as flash
crowds. Suppose $RE$ is the
service provider for a hotline. $SE_1$, $SE_2$
and $SE_3$ are three service providers that reach the hotline through $RE$. When
the hotline is activated, $RE$ is expected to receive a large call
volume to the hotline from $SE_1$, $SE_2$ and $SE_3$ that far
exceeds its usual call volume, potentially putting $RE$ into 
overload.

The second type of overload, known as UA-to-registrar overload, occurs  
when a large number of UAs overload their next hop server. A
typical example is avalanche restart, which happens when power is just
restored after a mass power failure in a large metropolitan area and a
huge number of SIP devices boot up trying to perform registration
simultaneously. This paper only discusses the proxy-to-proxy overload
problem. %

\subsection{TCP Window-based Flow Control Mechanism}

TCP is a reliable transport protocol with its built-in flow and
congestion control mechanisms. Flow control is exercised between two
TCP end points. The purpose of TCP flow control is to keep 
a sender from sending so much data that overflows the receiver's socket
buffer. %
Flow control is achieved by having the TCP receiver impose a receive
window on the sender side indicating how much data the receiver is willing
to accept at that moment; on the other hand, congestion control is the
process of a TCP sender imposing a congestion window by itself to avoid congestion inside the network. 
Thus, a TCP sender is governed by both the
receiver flow control window and sender congestion control window during its
operation. %

\begin{figure}
\centering
\epsfig{file=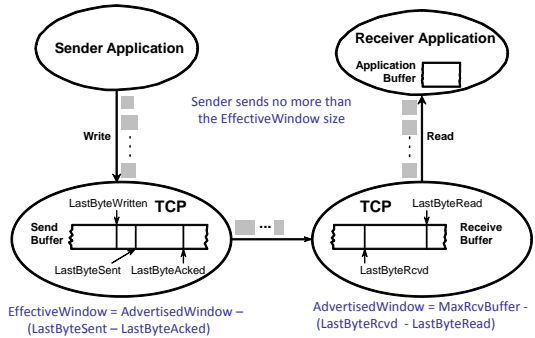, scale=0.3}
\caption{TCP flow control}
\label{fig:tcpcon}
\end{figure}

The focus of our work is on using TCP flow control since we are
interested in the receiving end point being able to deliver transport
layer feedback to the sending end point and we want to see how it could
facilitate higher layer overload control. We illustrate the
TCP flow control architecture in Fig.~\ref{fig:tcpcon}. A socket level
TCP connection usually maintains a send buffer and a receive buffer at
the two connection end points. The receiver application reads data from
the receive buffer to its application buffer.  The TCP 
receiver computes its current receive buffer availability as its
advertised window to the TCP sender. The TCP sender never sends more
data than an effective window size derived based on the receiver advertised window and data that
has been sent but not yet acknowledged.

\section{Experimental Testbed and Metrics}
\label{sec:testbed}

\subsection{Server and Client Software}
\label{sec-testbed-server}

We evaluated the Open SIP Server (OpenSIPS) version 1.4.2~\cite{opensips}, a freely-available, open source SIP proxy server. OpenSIPS is a fork of OpenSER, which in turn is a fork of the SIP Express Router (SER)\cite{ser}. These sets of servers represent the {\em de facto} open source version of SIP server, occupying a role similar to that of Apache for web servers. 
We also implemented our overload control mechanisms on the OpenSIPS server.      

We choose the widely used open source tool, SIPp~\cite{sipp} (May 28th 2009 release) to generate SIP traffic. We also make corrections to SIPp for our test cases. For example, we found that the SIPp does not trigger the {\sf 200 OK} retransmission timer over TCP as required by the SIP specification, and therefore we added it.

\subsection{Hardware, Connectivity and OS}
\label{sec:hardware}

The overloaded SIP RE server has $2$ Intel Xeon $3.06$\,GHz processors with $4$\,GB RAM. However, for our experiments, we only use one processor because SIP performance under multiple processors or a multi-core processor is itself a topic that requires separate attention~\cite{wright09:multicore}. We use up to $10$ machines for SEs, and up to $10$ machines for UACs. All the SE and UAC machines either have 2 Intel Pentium 4 $3.00$\,GHz processors with $1$\,GB memory or 2 Intel Xeon $3.06$\,GHz processors and $4$\,GB RAM. The server and client machines communicate over copper Gigabit or $100$\,Mbit Ethernet. The round trip time measured by the \texttt{ping} command between the machines is around $0.2$\,ms. More constrained link transmission conditions such as longer delays or explicit packet losses may be considered in future experiments.   

All of our testbed machines run Ubuntu 8.04 with Linux kernel 2.6.24. The default TCP send buffer size is 16\,KB and the default TCP receive buffer size is 85\,KB. Since the Linux operating system uses about 1/4 of the socket receive buffer size for bookkeeping overhead, the estimated effective default receive buffer size is about 64\,KB. In the rest of the paper we use the effective value to refer to receive buffer sizes. The SIP server application that we use allocates a default 64\,KB application buffer.  

\begin{cf-only}

Linux provides the {\tt setsockopt} API call to allow applications to manipulate connection-specific send and receive socket buffer sizes.  Linux also supports API calls that enable the applications to retrieve real-time status information about the underlying TCP connection. For example, using the {\tt ioctl} call, the application can learn about the amount of unsent data currently in the socket send buffer.  

\end{cf-only}

\subsection{Test Suite, Load Pattern and Performance Metrics}
\label{sec-workload}

We wrote a suite of Perl and Bash scripts to automate running the experiments and analyzing results. Our test load pattern is the same as in Fig~\ref{fig:sip-call-flow}. For simplicity but without loss of generality, we do not include call holding time and media. That means, the UAC sends a {\sf BYE} request immediately after sending an {\sf ACK} request. In addition, we do not consider the time between the ringing and the actual pick-up of the phone. Therefore, the UAS sends a {\sf 200 OK} response immediately after sending a {\sf 180 Ringing} response. In order to facilitate the load generation for overload tests, we also introduced extra cryptographic functions to the authentication operations in the SIP sessions to contrain the default server capacity.   
 
Our main performance metrics is the server throughput, i.e., number of sessions successfully set up per-second by receiving the {\sf ACK} to {\sf 200 OK} at UAS. We also examine a delay metrics similar to the Post Dial Delay (PDD) in PSTN networks, which roughly corresponds to the time from sending the first {\sf INVITE} to receiving the {\sf 200 OK} response. The combination of both throughput and delay metrics actually gives us the system goodput. A number of other metrics such as CPU utilization and server internal message processing rate are also used in explaining the results.

\section{Default SIP Over TCP Overload Performance}
\label{sec:results-default}

\begin{figure} [thbp]
\centering
\epsfig{file=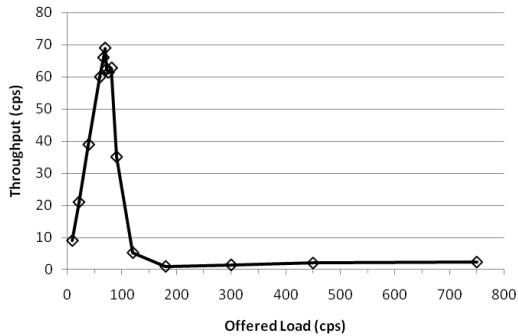,scale=0.6} 
\caption{Default SIP-over-TCP throughput} 
\label{fig:default-throughput}
\end{figure}

We start our investigation with a single SE - single RE testbed with all out-of-the-box configurations. The SE is connected to a machine acting as many UACs that generate the desired rate of SIP requests; the RE is connected to a machine acting as many UASes that receive and process SIP requests. The throughput results in calls per second (cps) of this testbed are shown in Fig.~\ref{fig:default-throughput}. It can be seen that the throughput immediately collapses as the load approaches and exceeds the server capacity at around 65 to 70\,cps. In this section, we explore the detailed causes of this behavior through server instrumentation.  

\begin{figure*}[thbp]
  \begin{center}
      \subfigure[{\sf INVITE}]{\label{fig:default-re-inv}
          \includegraphics[scale=0.6]{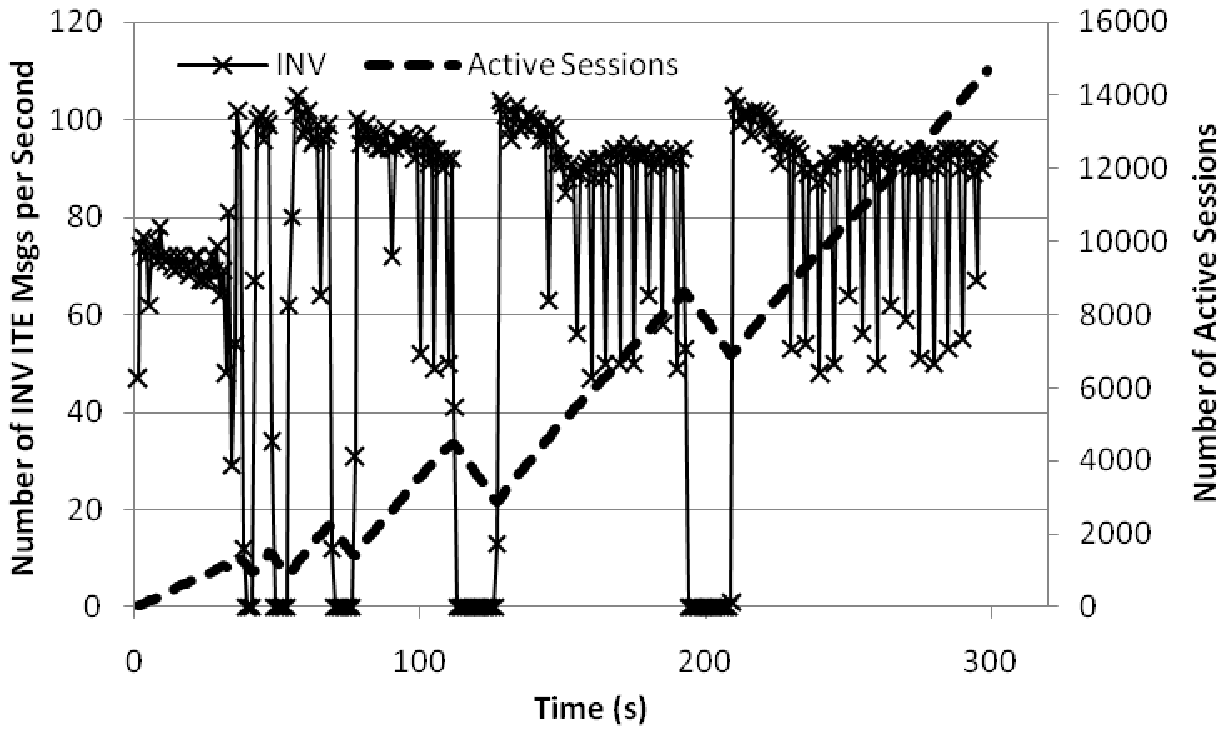}}
      \subfigure[{\sf BYE}]{\label{fig:default-re-bye}
        \includegraphics[scale=0.6]{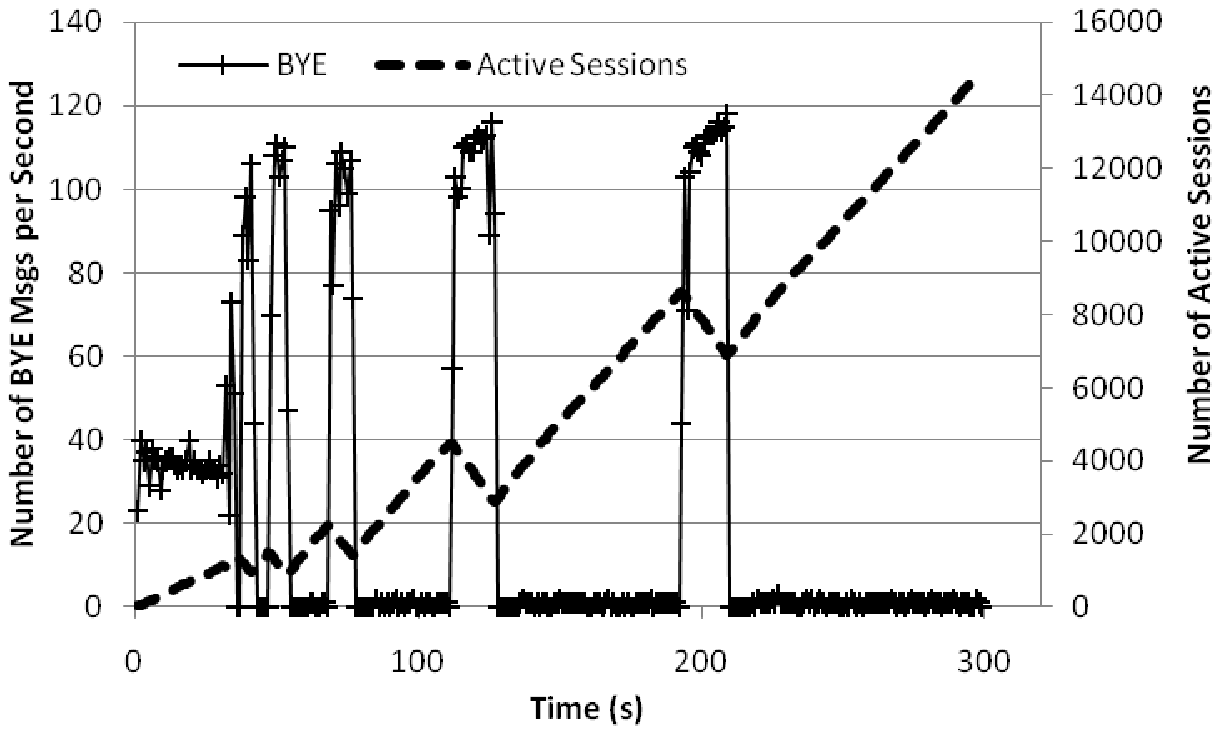}}
      \subfigure[{\sf 200 OK}]{\label{fig:default-re-200}
          \includegraphics[scale=0.6]{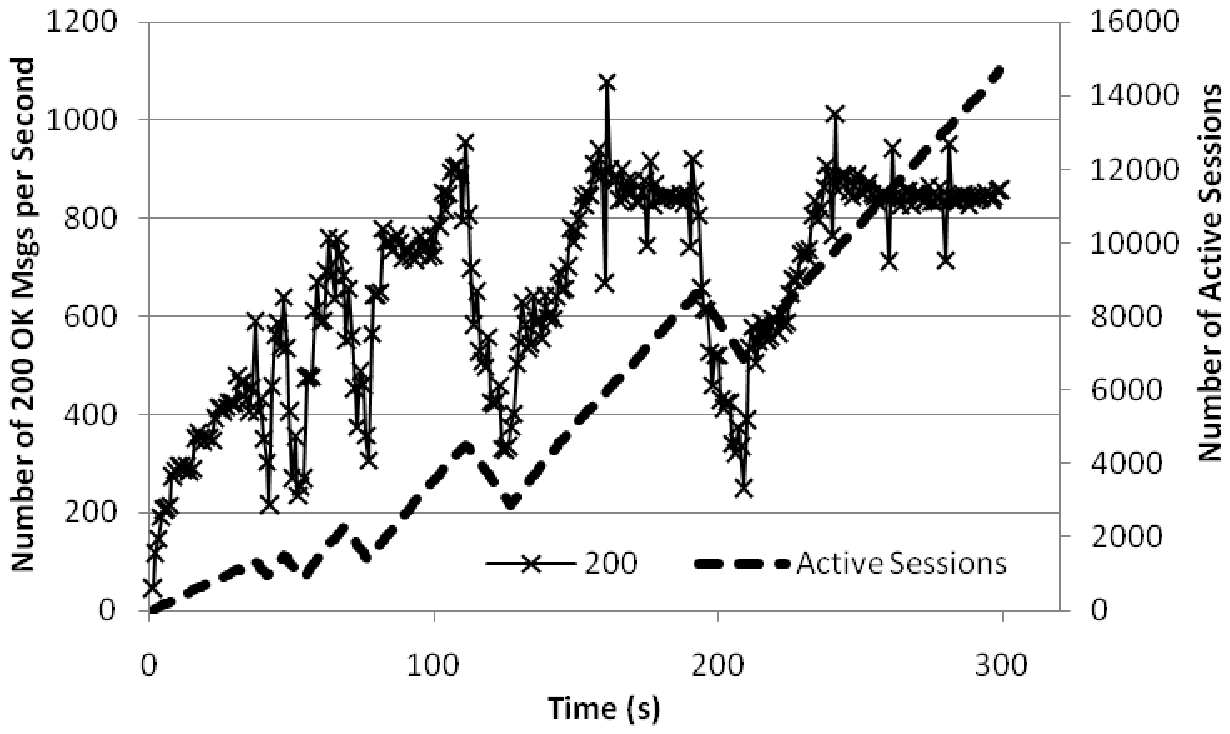}}
      \subfigure[{\sf ACK}]{\label{fig:default-re-ack}
        \includegraphics[scale=0.6]{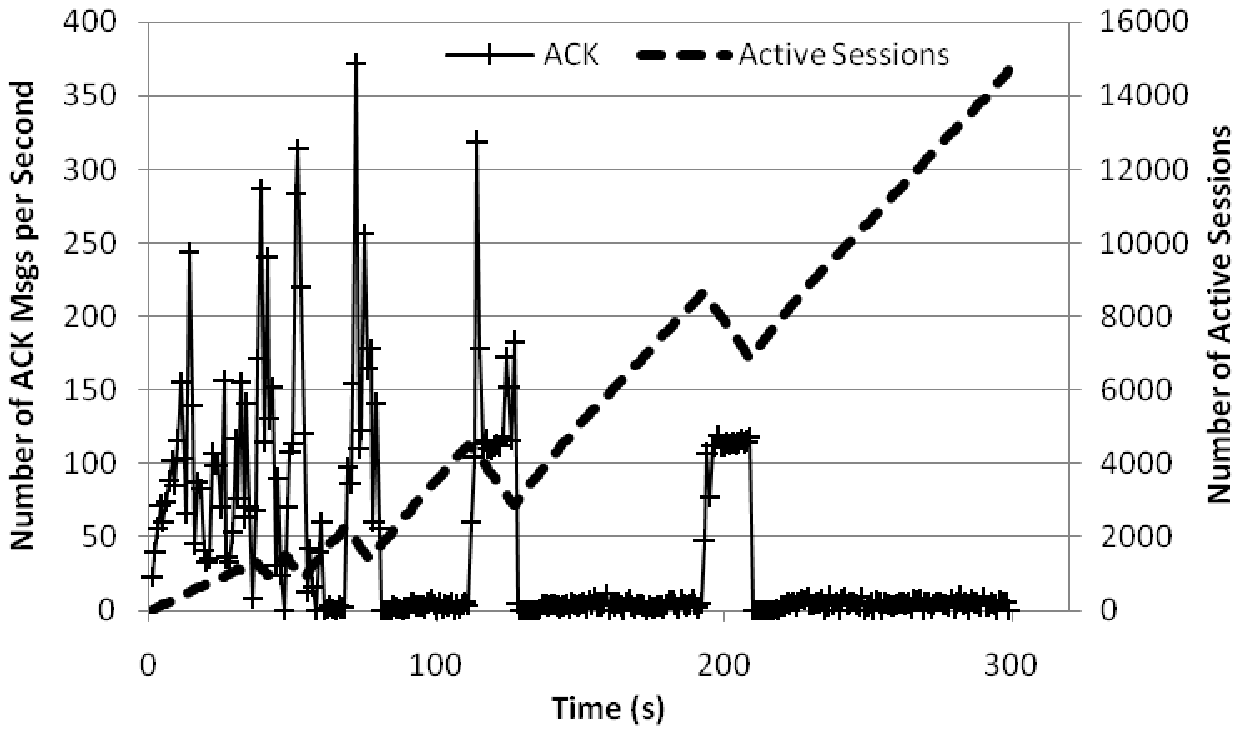}}
    \caption{RE message processing rates and number of active sessions in default SIP-over-TCP test} \label{fig:RE-default}
  \end{center}
\end{figure*}

We examine a particular run at a load of 150\,cps which is about 2.5 times the server capacity. Fig.~\ref{fig:RE-default} depicts the per second message processing rate. The four figures show {\sf INVITE}, {\sf BYE}, {\sf 200 OK} and {\sf ACK}, respectively. It should be noted that the number of {\sf 180 Ringing}s, not shown in these figures, basically follows the number of {\sf INVITE}s processed, because the UAS is not overloaded and can always deliver responses to RE. For the same reason, the number of {\sf 200 OK}s to {\sf BYE}s which are also not shown, follows the number of {\sf BYE}s. Along with the individual message processing rates, Fig.~\ref{fig:RE-default} also includes the current number of active sessions in the RE. The active sessions are those sessions that have been started by an {\sf INVITE} but have not yet received a {\sf BYE}. Since the call holding time is zero, in an ideal situation, any started sessions should be terminated immediately, leaving no session outstanding in the system. In a real system, the number of active sessions could be greater than zero. The larger the number of such in-progress sessions, the longer the delay that those sessions will experience. 

Fig.~\ref{fig:RE-default} indicates that {\sf 200 OK} retransmission happens almost immediately as the test starts, which means the end-to-end round trip delay immediately exceeds 500\,ms. This is caused by the large buffers at the different stages of the network system, which allow too many sessions to be accepted. The SIP session load is not atomic. The {\sf INVITE} request is always first introduced into the system and then come the responses and follow-up {\sf ACK} and {\sf BYE} requests. When too many {\sf INVITE}s are admitted to the system, the {\sf BYE} generation rate cannot keep up with the {\sf INVITE}s, resulting in a large number of active sessions in the system and also a large number of messages queued in various stages of the buffers. These situations translate to prolonged delays in getting the {\sf ACK} to {\sf 200 OK} to the UAS. More specifically, assuming the server's capacity is 65\,cps, if the sessions are indeed atomic, each session will take a processing time of 15.4\,ms. In order to avoid {\sf 200 OK} retransmission, the end-to-end one-way delay cannot exceed 250\,ms, corresponding to a maximum of about 16 active sessions in the system. Factoring in the non-atomic nature of the session load, this maximum limit could be roughly doubled to 32. But with the default system configuration, we have a 16\,KB TCP socket send buffer, and 64\,KB socket receive buffer, as well as 64\,KB SIP server application buffer. Considering an {\sf INVITE} size of around 1\,KB, this configuration means the RE can be filled with up to 130 {\sf INVITE}s at one time, much larger than the threshold of 32. All these {\sf INVITE}s contribute to active sessions once admitted. In the experiment, we see the number of active sessions reaches 49 at second 2, immediately causing {\sf 200 OK} retransmissions. 
{\sf 200 OK} retransmissions also trigger re-generated {\sf ACK}s, adding more traffic to the network. This is why during the first half of the time period in Fig.~\ref{fig:RE-default}, the number of {\sf ACK}s processed is higher than the number of {\sf INVITE}s and {\sf BYE}s processed. Eventually the RE has accumulated too many {\sf INVITE}s both in its receive buffer and application buffer. So its flow control mechanism starts to advertise a zero window to the SE, blocking the SE from sending additional {\sf INVITE} requests. Subsequently the SE stops processing {\sf INVITE} requests because of the send block to the RE. This causes SE's own TCP socket receive buffer and send buffer to get full as well. The SE's flow control mechanism then starts to advertise a zero window to UAC. This back pressure on UAC prevents the UAC from sending anything out to the SE. Specifically, the UAC can neither generate new {\sf INVITE} requests, nor generate more {\sf ACK} and {\sf BYE}s, but it could still receive responses. When this situation happens, retransmitted {\sf 200 OK}s received can no longer trigger retransmitted {\sf ACK}s. Therefore, the number of {\sf ACK}s processed in the later half of the graph does not exceed the number of {\sf INVITE}s or {\sf BYE}s. The number of {\sf ACK}s actually becomes similar to the number of {\sf BYE}s because {\sf BYE}s and {\sf ACK}s are generated together at the same time in our workload.  

It can further be seen that under the default settings, the {\sf INVITE} and {\sf BYE} processing tends to alternate with gradually increasing periods as the test proceeds. During each period, the {\sf INVITE} portion is increasingly larger than the {\sf BYE} portion. Since the number of active sessions always increases with {\sf INVITE} processing, and decreases with {\sf BYE} processing, those processing patterns lead to the continued growth of the number of active sessions in the RE and exacerbate the situation.

\begin{figure} [thbp]
\centering
\epsfig{file=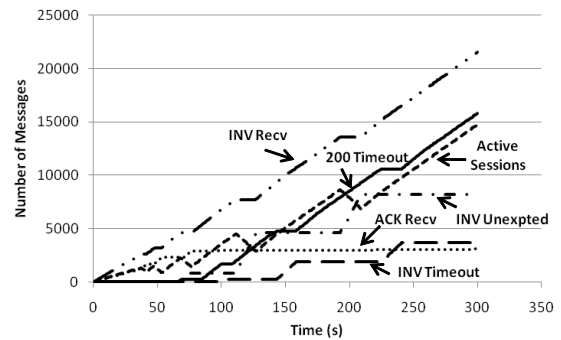,scale=0.6} 
\caption{Total number of messages processed at UAS and number of active sessions at RE} 
\label{fig:default-uas-count}
\end{figure}

In addition to observing the per-second message processing rate at RE, we also confirm the behavior from the total number of messages processed at the UAS, along with the number of active sessions at RE as in Fig.~\ref{fig:default-uas-count}. Note that the numbers of {\sf INVITE}s received, {\sf 180 Ringing} and initial {\sf 200 OK} (not retransmissions) messages sent are the same, because {\sf 180 Ringing} and {\sf 200 OK} are generated by UAS immediately upon receiving an {\sf INVITE}. Similarly the number of {\sf ACK}, {\sf BYE}, and {\sf 200 OK} to {\sf BYE}s are the same, because {\sf ACK} and {\sf BYE} are generated at the same time at the UAC and {\sf 200 OK} to {\sf BYE} is immediately generated upon receiving {\sf BYE} at the UAS. In Fig.~\ref{fig:default-uas-count}, initially between 0 and the 38th second, the numbers of {\sf ACK}s and {\sf BYE}s received are roughly half of the total {\sf INVITE}s received. Therefore, the number of active sessions in the RE and the number of {\sf ACK}s received at the UAS are roughly the same. Then RE enters the abnormal {\sf INVITE} processing and {\sf BYE} processing alternating cycle. During the period when RE is processing {\sf ACK}s and {\sf BYE}s, the number of active sessions decreases. During the period when RE is processing {\sf INVITE}s, no {\sf ACK}s are forwarded, so the number of {\sf ACK}s remains constant. 

{\sf 200 OK} retransmission starts at second 2. The total period of {\sf 200 OK} retransmission lasts 32 seconds for each individual session, therefore the expiration of the first session that has exhausted all its {\sf 200 OK} retransmissions without receiving an {\sf ACK} happens at the 34th second. The actual {\sf 200 OK} retransmission timeout we see from Fig.~\ref{fig:default-uas-count} is at the 66th second. The difference between the 66th and 34th second is 32 seconds, which is a configured maximum period UAS waits to receive the next message in sequence, in this case the {\sf ACK} corresponding to the {\sf 200 OK}.    

Starting from the 69th second, we see a category of messages called {\em INVITE Unexpected}. These are {\sf ACK}s and {\sf BYE}s that arrive after the admitted sessions have already timed out at the UAS. These {\sf ACK}s and {\sf BYE}s without a matching session also create session states at the SIPp UAS, which normally expect a session message sequence beginning with an {\sf INVITE}. Since those session states will not receive other normal in-session messages, at the 101th second, or after 32\,seconds of UAS receive timeout period, those session states start to time out, reflected in the figure as the {\em INVITE Timeout} curve. Finally, a very important overall observation from Fig.~\ref{fig:default-uas-count} is that at a certain point, the 77th second, the number of timely received {\sf ACK}s virtually stopped growing, causing the throughput to drop to zero. 

\begin{figure}[thbp]
  \begin{center}
      \subfigure[UAC]{\label{fig:default-devp-uac-screen}
          \includegraphics[scale=0.43]{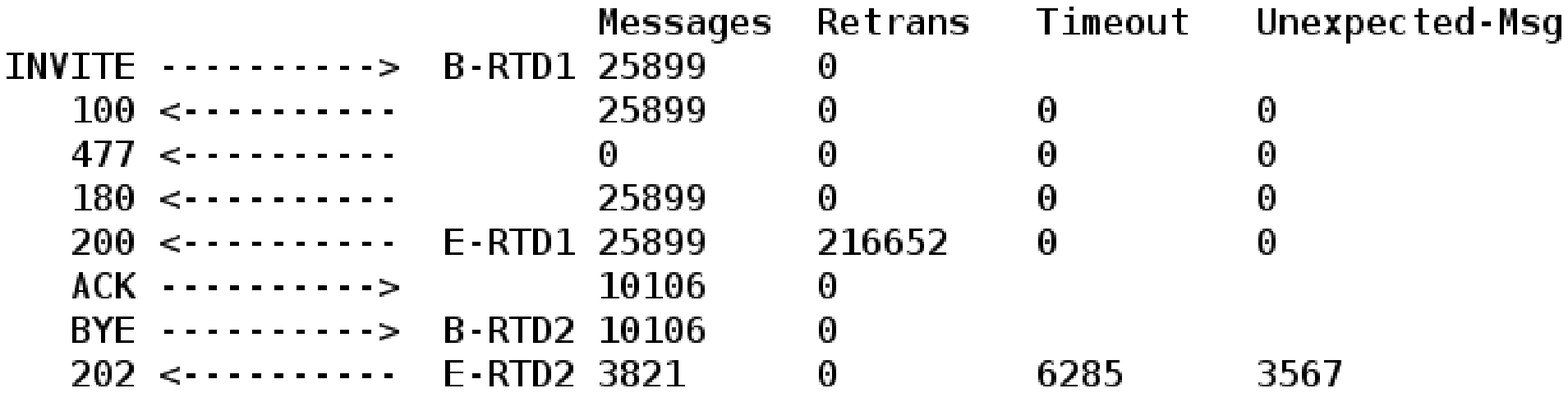}}
      \subfigure[UAS]{\label{fig:default-devp-uas-screen}
        \includegraphics[scale=0.41]{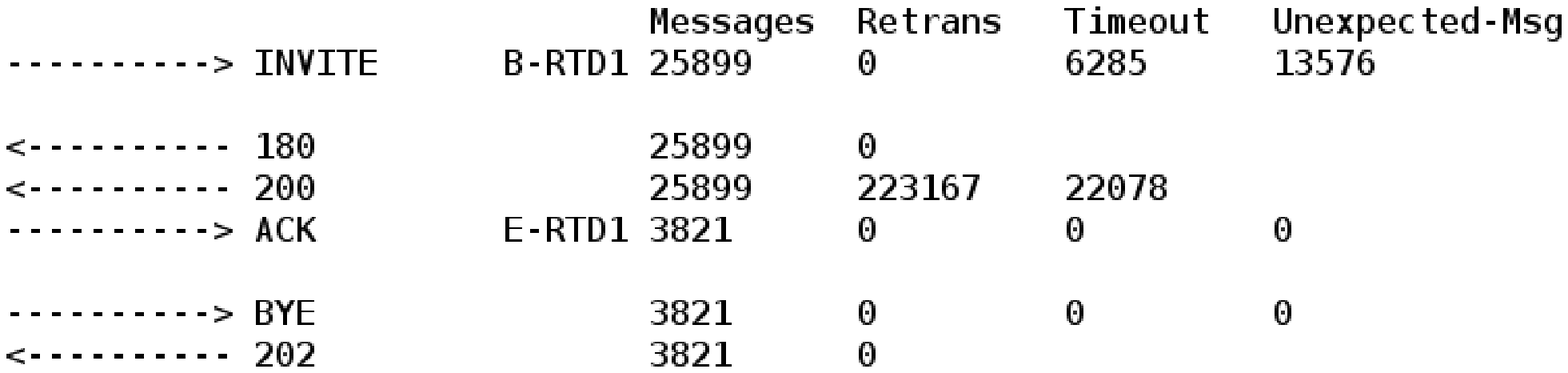}}
    \caption{Screen logs in default SIP-over-TCP test}
    \label{fig:default-screen}
  \end{center}
\end{figure}

We also show the final screen logs at the UAC and UAS sides for the test with default configurations in Fig.~\ref{fig:default-screen}, where status code {\sf 202} is used instead of {\sf 200} to differentiate the {\sf 200 OK} to {\sf BYE} from the {\sf 200 OK} to {\sf INVITE}. Earlier in this section we have explained the {\sf 200 OK} retransmissions, {\sf 200 OK} timeouts, {\sf INVITE} timeouts, and {\sf INVITE}s unexpected messages. We can see that among the 25,899 {\sf INVITE}s received at the UAS side, 22,078 eventually time out and only 3,821 receive the final {\sf ACK}. The UAC actually sends out a total of 10,106 {\sf ACK}s and {\sf BYE}s. The remaining 6,285 {\sf ACK}s and {\sf BYE}s are eventually delivered to UAS but are too late when they arrive, therefore those {\sf BYE}s do not trigger {\sf 202 OK} and we see 6,285 {\sf 202 OK} timeouts at the UAC. At the UAS side, those 6,285 {\sf ACK}s and {\sf BYE}s establish abnormal session states and eventually time out after the 32\,s receive timeout for {\sf INVITE}. The unexpected messages at the UAC side are {\sf 408 Send Timeout} messages triggered at the SIP servers for the {\sf BYE}s that do not hear a {\sf 202 OK} back. Note that the number of those messages (3,567) is smaller than the exact number of {\sf BYE}s that do not receive {\sf 202 OK} (6,285). This is because the remaining 2,718 {\sf 408 Send Timeout} messages arrive after the {\sf 202 OK} receive timeout and therefore those messages were simply discarded and not counted in the screen log.   

\begin{cf-only}

Finally, we also measure the PDD and find that even without considering whether {\sf ACK}s are delivered successfully, 73\% of the {\sf INVITE}s have PDDs between 8 and 16 seconds, which are most likely beyond the human interface acceptability limit. Another 24\% have PDDs between 4 to 8 seconds, which might be close to the acceptable limit. 

\end{cf-only}

\section{SIP-over-TCP Overload Control Mechanism Design} \label{sec:ourmec}

From the SIP-over-TCP congestion collapse, we learned a key lesson that we must limit the number of {\sf INVITE}s we can admit to avoid too many active sessions accumulating in the system. For all admitted {\sf INVITE}s, we need to make sure the rest of the session messages complete within finite delay. In this section, we propose specific approaches to address these issues, namely {\em connection split}, {\em buffer minimization}, and {\em smart forwarding}.

\subsection{Connection Split and Buffer Minimization}

\begin{figure} [thbp]
\centering
\epsfig{file=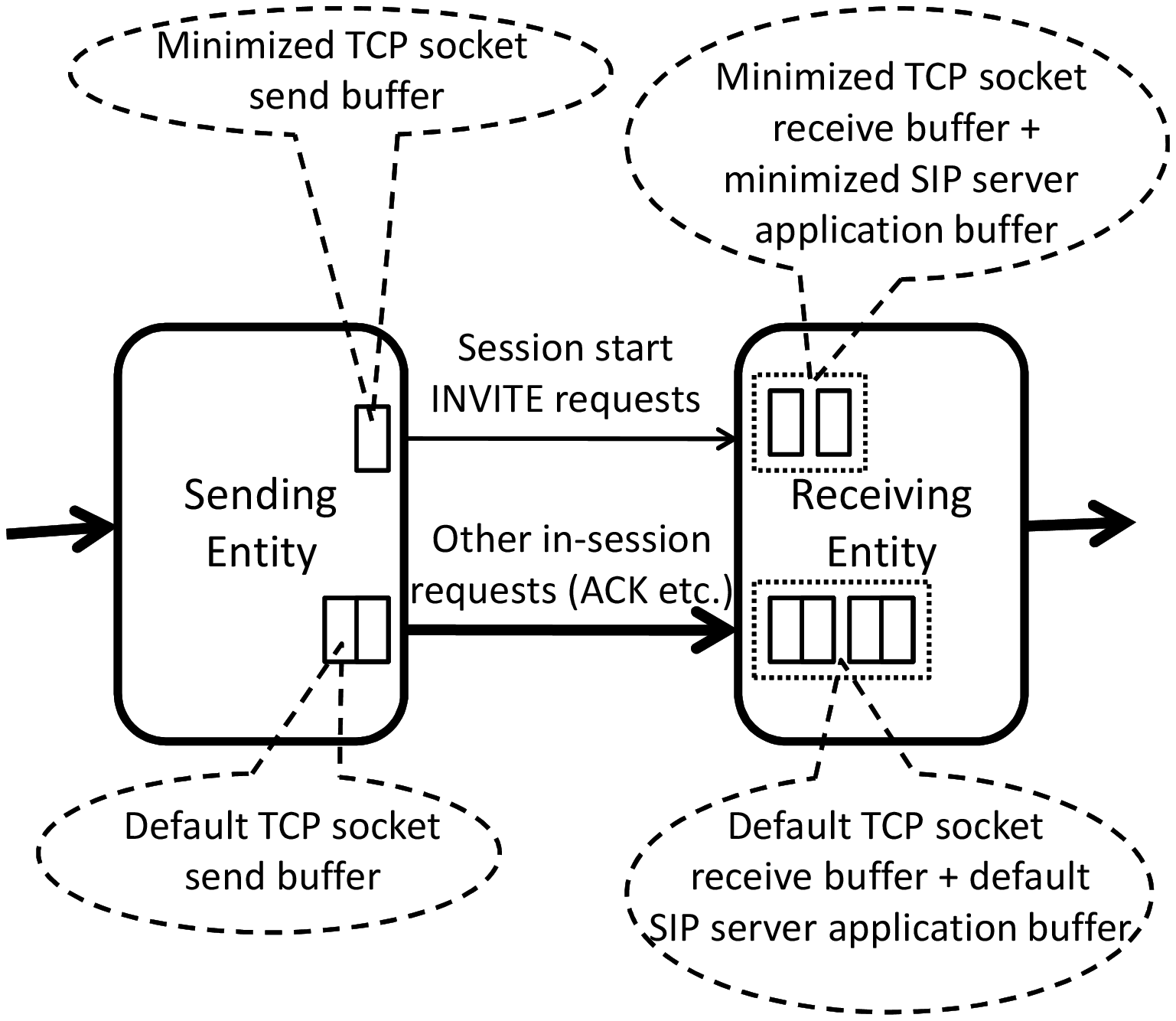,scale=0.4} 
\caption{ECS + BM} \label{fig:dc-nosf}
\end{figure}

First, it is clear that we only want to limit {\sf INVITE}s but not non-{\sf INVITE}s because we do not want to drop messages for sessions already accepted. In order to have a separate control of {\sf INVITE}s and non-{\sf INVITE} messages, we split the TCP connection from SE to RE into two, one for {\sf INVITE} requests, and the other for all other requests. In other words, the RE will listen on two TCP connections, and the SE makes sure that it will send all {\sf INVITE}s to one connection and all non-{\sf INVITE}s to the other connection. Second, in order to limit the number of {\sf INVITE}s in the system and minimize delay, we minimize the total system buffer size between the SE and the RE for the {\sf INVITE} connection, which includes three parts: the SE TCP socket send buffer, the RE TCP socket receive buffer and the RE SIP server application buffer. We call the resulting mechanism {\em Explicit Connection Split + Buffer Minimization} (ECS+BM) and illustrate it in Fig.~\ref{fig:dc-nosf}.   

We find, however, although ECS+BM effectively limits the number of {\sf INVITE}s that could accumulate at the RE, the resulting throughput differs not much from that of the default configuration. The reason is that, since the number of {\sf INVITE}s SE receives from UAC remains the same and the {\sf INVITE} buffer sizes between SE and RE are minimized, the {\sf INVITE} pressure merely moves a stage back and accumulates at the UAC-facing buffers of the SE. Once those buffers, including the SE receive buffer and SE SIP server application buffer, have been quickly filled up, the system delay dramatically increases. Furthermore, the UAC is then blocked from sending to SE and unable to generate {\sf ACK}s and {\sf BYE}s, causing the number of active sessions in the RE to skyrocket. In conclusion, ECS+BM by itself is insufficient in preventing overload. 

\begin{figure} [thbp]
\centering
\epsfig{file=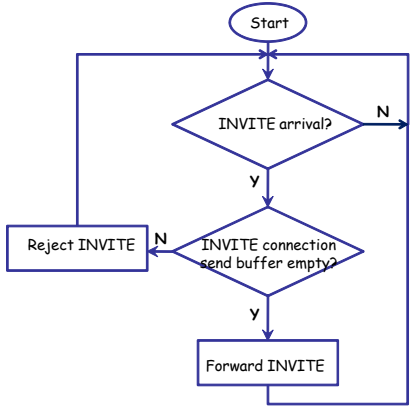,scale=0.4} 
\caption{Smart forwarding for ECS} \label{fig:sf4dc}
\end{figure}
 
\subsection{Smart Forwarding}

In order to release, rather than pushing back the excessive load pressure present in the ECS+BM mechanism, we introduce the {\em Smart Forwarding} (SF) algorithm as shown in Fig.~\ref{fig:sf4dc}. This algorithm is enforced only for the {\sf INVITE} connection. When an {\sf INVITE} arrives, the system checks whether the current {\sf INVITE} connection send buffer is empty. If yes, the {\sf INVITE} is forwarded; otherwise the {\sf INVITE} is rejected with an explicit SIP rejection message. This algorithm has two advantages: first, although we can choose any send buffer length threshold value for rejecting an {\sf INVITE}, the decision to use the emptiness criterion makes the algorithm parameter-free; second, implementation of this algorithm is especially easy in Linux systems because the current send buffer occupancy can be retrieved by a simple standard \texttt{ioctl} call. 

\begin{cf-only}
Our resulting mechanism is then ECS+BM+SF. We evaluate its performance on our testbed from light to heavy overload and find it achieving nearly full system capacity all the time. %
Due to space limitation, we do not present the results of the ECS+BM+SF here, but discuss in more detail an even simpler mechanism developed based on it called ICS+BM+SF.  
\end{cf-only}

\begin{figure} [thbp]
\centering
\epsfig{file=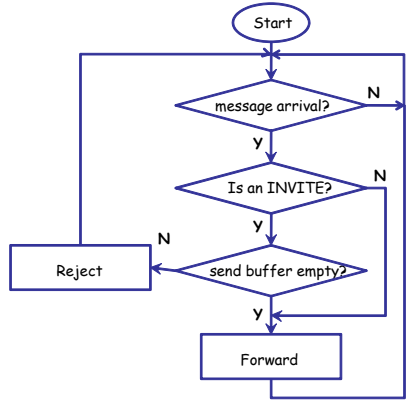,scale=0.4} 
\caption{Smart forwarding for ICS} \label{fig:sf4vdc}
\end{figure}

\subsection{Implicit Connection Split, Buffer Minimization and Smart Forwarding (ICS+BM+SF)} \label{sec:vcs+bm+sf}

\begin{cf-only}
Our results show that the ECS+BM+SF mechanism is very effective. Even in high overload, the RE contains only a few active sessions all the time, and achieves full capacity. The only inconvenience is that it requires to establish two separate connections for {\sf INVITE}s and non-{\sf INVITE}s. But if the server is never backlogged, the queue size for both {\sf INVITE} and non-{\sf INVITE} request connections should be close to zero. In that case, the dedicated connection for non-{\sf INVITE} requests does not require the default large buffer setting either. We therefore decide to merge the two split connections back into one but still keep the minimized SE send buffer, RE receive buffer and application buffer settings. We also need to revise our {\em smart forwarding} algorithm accordingly, as in Fig.~\ref{fig:sf4vdc}. Since there is only a single request connection now, the algorithm performs an additional check for {\sf INVITE} requests and rejects it if the send buffer is non-empty. Otherwise, the {\sf INVITE} is forwarded. All non-{\sf INVITE} requests are always forwarded. Although the revised mechanism no longer requires a dedicated connection for {\sf INVITE}s, it treats {\sf INVITE}s and non-{\sf INVITE}s differently. Therefore, we call this revised mechanism {\em Implicit Connection Split} (ICS) as opposed to the previous ECS mechanism. %
\end{cf-only}

\begin{figure} [thbp]
\centering
\epsfig{file=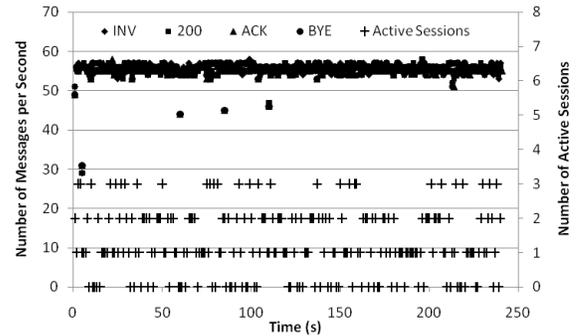,scale=0.6} 
\caption{RE message processing rates with ICS+MB+SF} \label{fig:sc+mb+sf-re-count}
\end{figure}

\begin{figure}[thbp]
  \begin{center}
      \subfigure[UAC]{\label{fig:sc+mb+sf-devp-uac-screen}
          \includegraphics[scale=0.45]{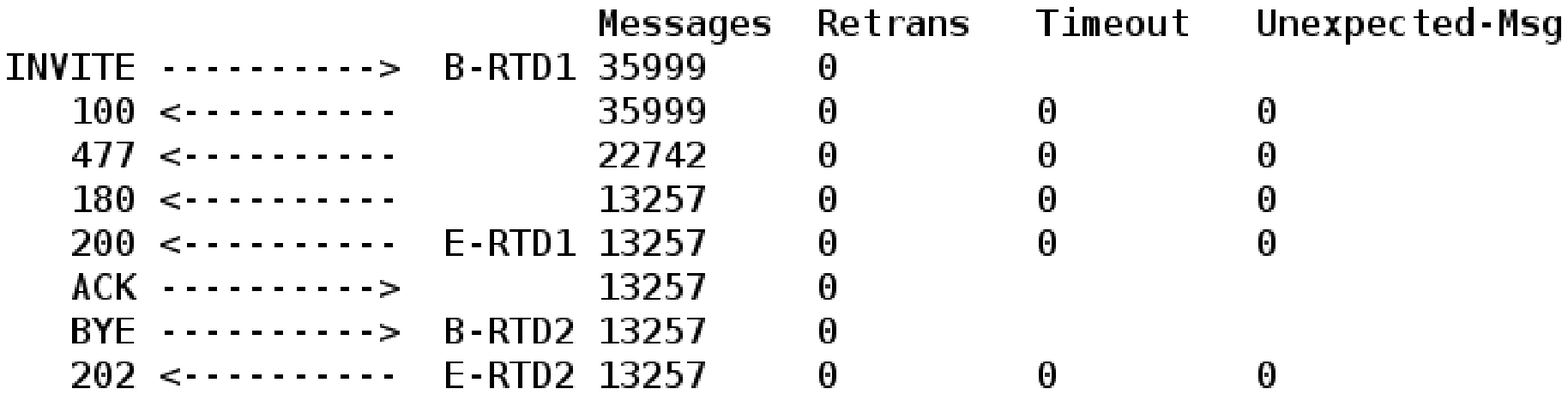}}
      \subfigure[UAS]{\label{fig:sc+mb+sf-devp-uas-screen}
        \includegraphics[scale=0.42]{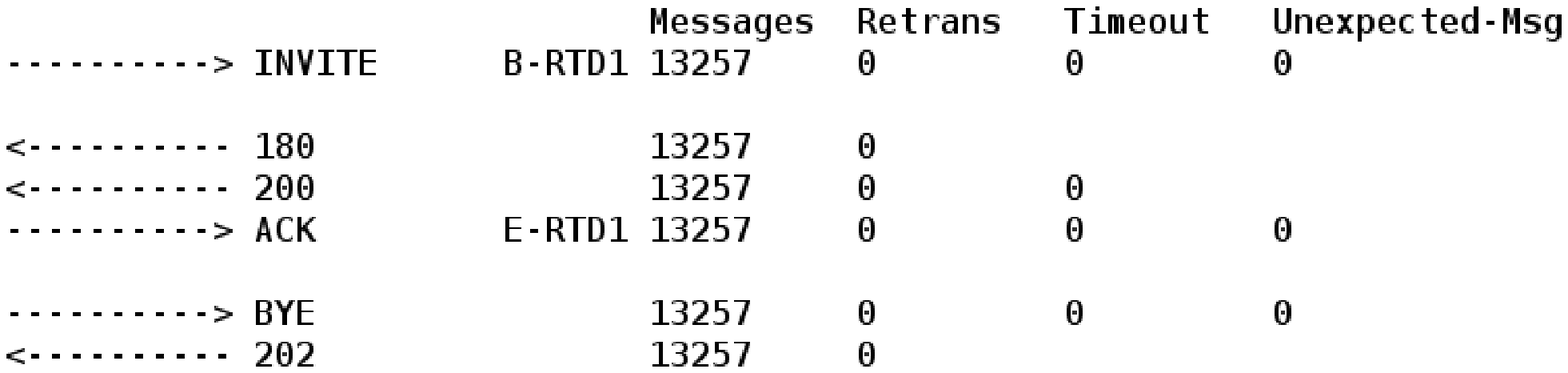}}
    \caption{Screen logs with ICS+MB+SF}
    \label{fig:sc-screen}
  \end{center}
\end{figure}

\begin{cf-only}
We evaluate the resulting ICS+BM+SF mechanism and compare its performance with the default configuration in the same scenario as in Section~\ref{sec:results-default} with one SE overloading an RE at an offered load of 2.5\,times the server capacity. Fig.~\ref{fig:sc+mb+sf-re-count} shows the average message processing rate and the number of active sessions in the RE. We can see how this figure differs dramatically from Fig.~\ref{fig:RE-default}. Here, the values of {\sf INVITE}, {\sf 200 OK}, {\sf ACK}, and {\sf BYE} processing rate overlap most of the time, which explains why the number of active sessions remains extremely low, between 0 and 3, all the time. Furthermore, from the overall UAC and UAS screen logs in Fig.~\ref{fig:sc-screen}, we see that among the 35,999 {\sf INVITE}s that are generated, 22,742 of them are rejected by the {\em smart forwarding} algorithm. The remaining 13,257 sessions all successfully get through, without triggering any retransmission or unexpected messages - a sharp contrast to Fig.~\ref{fig:default-screen}. The good performance is also shown by the PDDs. We find that over 99.8\% of the sessions have a delay value smaller than 30\,ms, far smaller than the 500\,ms {\sf 200 OK} retransmission threshold. Finally, the system achieves full capacity as confirmed by the full CPU utilization observed at the RE. %

\end{cf-only}

\begin{cf-only}

\subsection{Parameter Tuning} \label{sec:subpmtune}

Our ICS+BM+SF mechanism in section~\ref{sec:vcs+bm+sf} contains three minimized buffer sizes: the SE send buffer at 2\,KB, RE receive buffer at 1\,KB and RE application buffer at 1,200\,bytes. We conducted extensive tests to explore the impact of tuning these three buffer sizes, and we summarize the results in this section.

First, we find that since the RE receive buffer and RE application buffer are connected in series, they do not have to be minimized at the same time. Minimizing either one of them achieves similar near-capacity throughput. However, recall that enlarging either RE buffer size could hold messages in the RE and increase queuing delay. For example, we plot the PDD distribution for four test cases in Fig.~\ref{fig:pdd-def-ab-rb}. Two of those cases compare the delay when the RE application buffer is set to 2\,KB vs. the default 64\,KB, while the RE receive buffer is at its default value of 64\,KB. Most of the delays in the small application buffer case are below 375\,ms, and as a result we observe no {\sf 200 OK} retransmissions at the UAS side. In the large application buffer case, however, nearly 70\% of the sessions experience a PDD between 8\,seconds and 32\,seconds, which will most likely be hung up by the caller even if the session setup messages could ultimately complete. Not surprisingly, we also see a large number of {\sf 200 OK} retransmissions in this case. 

\begin{figure} [hbp]
\centering
\epsfig{file=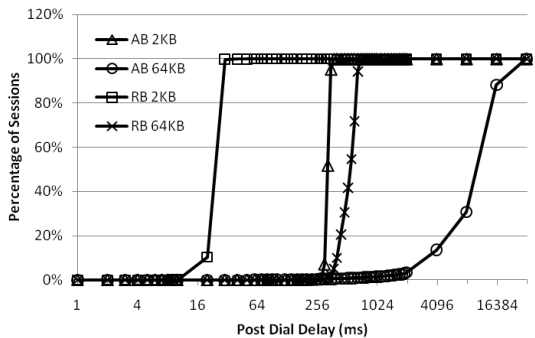,scale=0.6} 
\caption{PDD comparison for RE side buffer tuning (AB: Application Buffer; RB: Receive Buffer)} \label{fig:pdd-def-ab-rb}
\end{figure}

The other two cases in Fig.~\ref{fig:pdd-def-ab-rb} compare the PDD when the receive buffer is set to 2\,KB vs. the default 64\,KB, while the application buffer is at its default value of 64\,KB. In the small receive buffer case, over 99.7\% of the sessions have a PDD below 30\,ms, and there is certainly no {\sf 200 OK} retransmissions at the UAS side. In the larger receive buffer case, about 30\% of the sessions have a PDD below 480\,ms, and the remaining 70\% between 480\,ms and 700\,ms. Since a large number of sessions experienced a round trip delay exceeding 500\,ms, we see quite a number of {\sf 200 OK} retransmissions at the UAS side, too. Therefore, tuning the receive buffer is preferable over tuning the application buffer, which matches the intuition: the receive buffer is closer to the SE and produces more timely transport feedback than the application buffer does. 

Second, we find that the SE send buffer size actually does not have to be minimized. This can be attributed to our {\em smart forwarding} algorithm which already prevents excessive non-{\sf INVITE} messages from building up in the system. Combined with minimized buffers at the RE, our mechanism minimizes the number of active sessions in the system, which means there will always be only a small number of messages in the SE send buffer. 

In summary, our investigation confirms that the only essential tunable parameter of the ICS+BM+SF mechanism is the RE receive buffer size. Therefore, we finally obtain our extremely simple ICS+BM+SF mechanism as illustrated in Fig.~\ref{fig:mec-sim}. 

\begin{figure}[thbp]
\centering
\epsfig{file=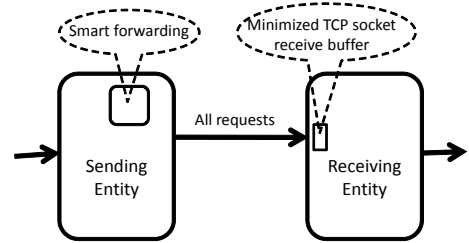,scale=0.4} 
\caption{ICS+BM+SF} \label{fig:mec-sim}
\end{figure}

\end{cf-only}

\section{Overall Performance of our SIP-over-TCP Overload Control Mechanisms} \label{sec:overallperf}

\begin{cf-only}
In this section we evaluate the overall performance of our ICS+BM+SF mechanism as shown in Fig.~\ref{fig:mec-sim}. To demonstrate scalability, we test on three scenarios with 1 SE, 3 SEs and 10 SEs, respectively.  
\end{cf-only}

\subsection{Overall Throughput and PDD}

\begin{cf-only}
\begin{figure} [thbp]
\centering
\epsfig{file=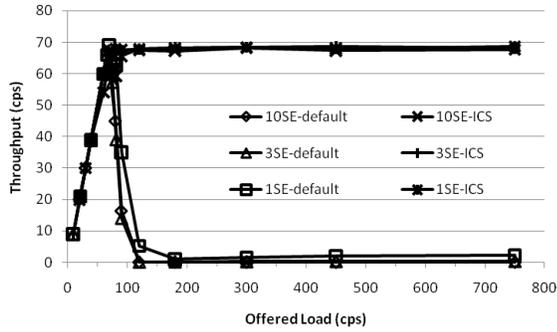,scale=0.6} 
\caption{Overall throughput of SIP-over-TCP: with and without our overload control mechanism} \label{fig:tp-overall}
\end{figure}
\end{cf-only}

\begin{figure} [htb]
\centering
\epsfig{file=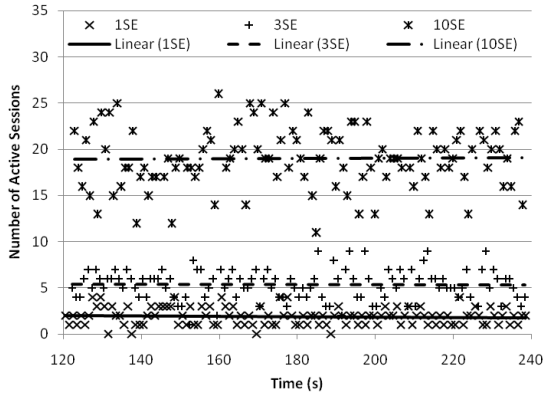,scale=0.6} 
\caption{Number of active sessions in RE in scenarios with varying number of SEs} \label{fig:ova-oscall}
\end{figure}

\begin{cf-only} Fig.~\ref{fig:tp-overall} illustrates the throughput with and without our control mechanism\end{cf-only} in the three test scenarios with varying number of SEs and an offered load up to over 10\,times the capacity. The RE receive buffer was set to 2\,KB and the SE send buffer and RE application buffer remain at their default values. As we can see, in all test runs with our control mechanisms, the overload throughput maintains at close to the server capacity, even in the most constrained case with 10\,SEs and a load of 750\,cps.
\begin{cf-only}
Moreover, we observe no single {\sf 200 OK} retransmissions in any of those tests. 
\end{cf-only}

\begin{cf-only}
We further compare the tests with different number of SEs.
\end{cf-only}

\begin{figure} [hbp]
\centering
\epsfig{file=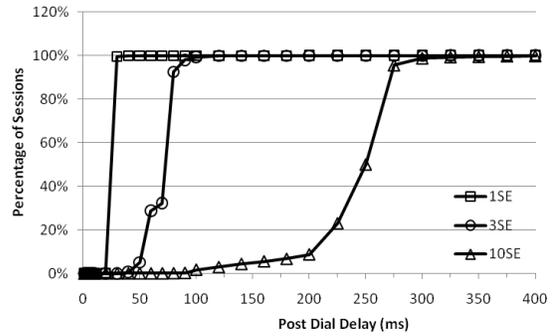,scale=0.6} 
\caption{PDD in scenarios with varying number of SEs} \label{fig:ova-pdd}
\end{figure}

Fig.~\ref{fig:ova-oscall} shows that the numbers of active sessions in RE for the three scenarios roughly correspond to the ratio of the numbers of SEs (1:3:10), as would be expected because in our testbed configuration each SE creates a new connection to the RE and is allocated a new set of RE buffers. Increased number of active sessions causes longer PDDs, as demonstrated in Fig.~\ref{fig:ova-pdd}, where the overall trend and the 50 percentile values match the 1:3:10 ratio pretty well.

\begin{cf-only}
Fig.~\ref{fig:ova-oscall} and Fig.~\ref{fig:ova-pdd} also imply that if the number of SEs keeps increasing, the system will eventually still accumulate an undesirably large number of active sessions. The PDD will also exceed the response retransmission timer value to cause {\sf 200 OK} retransmissions. 
\end{cf-only}

Thus, our mechanism is most applicable to cases where the number of SEs is reasonably small, which however, does cover a fairly common set of realistic SIP server overload scenarios. For example, there are typical national service providers deploying in total hundreds of core proxy and edge proxy servers in a hierarchical manner. The resulting server connection architecture leaves each server with a few to dozens of upstream servers.

\subsection{RE Receive Buffer Tuning}

\begin{figure} [thbp]
\centering
\epsfig{file=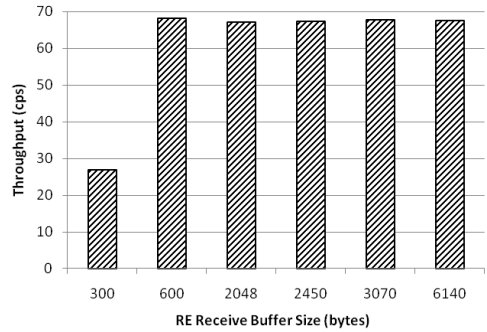,scale=0.6} 
\caption{Impact of RE receive buffer size on Throughput} \label{fig:tp-10se-rcvbuf}
\end{figure}

\begin{figure} [thbp]
\centering
\epsfig{file=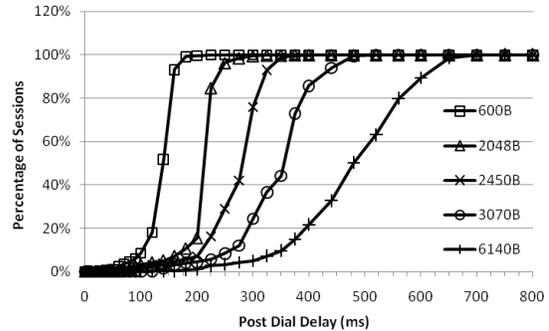,scale=0.6} 
\caption{Impact of RE receive buffer size on PDD} \label{fig:pdd-10se-rcvbuf}
\end{figure}

The only tunable parameter in our mechanism is the RE receive buffer size. We explore the impact of this parameter under the most constrained case where there are 10 SEs with a total load of 750\,cps  
in Fig.~\ref{fig:tp-10se-rcvbuf}. It is not surprising that the receive buffer size cannot be too small because that will cause a single message to be sent and read in multiple segments. After exceeding a certain threshold, the receive buffer does not make difference in overload throughput, but the smaller the buffer is, the lower the PDD, as shown in Fig.~\ref{fig:pdd-10se-rcvbuf}. The PDD is roughly the same as round trip delay. If the round trip delay exceeds 500\,ms, we will start to see {\sf 200 OK} retransmissions, as in the cases where the receive buffer is larger than 3,070\,bytes.

\begin{cf-only}
Overload control algorithms are meant to kick in when overload occurs. In practice, a desirable feature is to require no explicit threshold detection about when the overload control algorithm should be activated, because that always introduces additional complexity, delay and inaccuracy. If we keep our overload control mechanism on regardless of the load, then we should also consider how our mechanism could affect the system {\em underload} performance. We find that in general our mechanisms have a pretty satisfactory underload performance, meaning the throughput matches closely with a below-capacity offered load as shown in Fig.~\ref{fig:tp-overall}, although in some corner cases ICS's underload performance is not as good as ECS because ICS tends to be more conservative and reject more sessions. 
\end{cf-only}

Overall, in order to scale to as many SEs as possible yet minimizing the PDD, we recommend an RE receive buffer size that holds roughly a couple of {\sf INVITE}s.

\subsection{Fairness}

All our above tests with multiple SEs assume each SE receiving the same request rate from respective UACs, in which case the throughput for each UAC is the same. Now we look at the situation where each SE receives different request rates, and measure the fairness property of the achieved throughput.    

\begin{figure} [thbp]
\centering
\epsfig{file=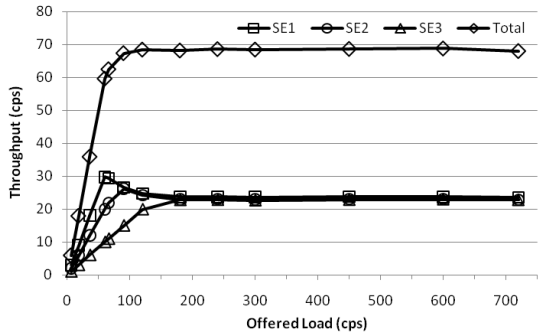,scale=0.6} 
\caption{Throughput: three SEs with incoming load ratio 3:2:1} \label{fig:tp-3se-fairness}
\end{figure}

Fig.~\ref{fig:tp-3se-fairness} shows the throughput of a 3\,SE configuration with the incoming offered load to the three SEs distributed at a 3:2:1 ratio. As we can see, when the load is below total system capacity, the individual throughputs via each SE follow the offered load at the same 3:2:1 ratio closely. At light to moderate overload until 300\,cps, the higher load sources have some advantages in competing RE resources. At higher overload above 300\,cps, each SE receives a load that is close to or higher than the server capacity. The advantages of the relatively higher load SEs are diminishing, and the three SEs basically deliver the same throughputs to their corresponding UACs. 

Shen {\em et al.}~\cite{shen08} define two types of fairness for SIP server overload: {\em service provider-centric} fairness and {\em end user-centric} fairness. The former allocates the same portion of the overloaded server capacity to each upstream server; the latter allocates the overloaded server capacity in proportion to the upstream servers' original incoming load. Our results show that the system achieves {\em service provider-centric} fairness at heavy overload. Obtaining {\em end user-centric} fairness during overload is usually more complicated; some related techniques are discussed in~\cite{shen08}.

\subsection{Additional Discussions}

During our work with OpenSIPS, we also discover subtle software implementation flaws or configuration guidelines. For example, an SE could block on sending to an overloaded RE. Thus, if there are new requests coming from the same server at the upstream of the SE but are destined to other REs that are not overloaded, those new requests cannot be accepted either. This head-of-line blocking effect is clearly a flaw that is hardly noticeable unless we conduct systematic TCP overload tests. 

Another issue is related to the OpenSIPS process configuration. OpenSIPS employs a multi-process architecture and the number of child processes is configurable. Earlier work~\cite{shen09techreport} with OpenSIPS has found that configuring one child process yields an equal or higher maximum throughput than configuring multiple child processes. However, in this study we find that when overloaded, the existing OpenSIPS implementation running over TCP with a single child process configuration could lead to a deadlock between the SE and RE servers. Therefore, we use multiple child processes for this study.  

\section{Conclusions} \label{sec:conclude}

We experimentally evaluated default SIP-over-TCP overload performance
using a popular open source SIP server implementation on a typical
Intel-based Linux testbed. Through server instrumentation, we found
that TCP flow control feedback cannot prevent SIP overload congestion collapse because of lack of
application context awareness at the transport layer for session-based
load with real-time requirements. We develop novel mechanisms that
effectively use existing TCP flow control to aid SIP application level overload
control. Our mechanism has three components: the first is {\em connection
split} which brings a degree of application level
awareness to the transport layer; the second is a parameter-free {\em
smart forwarding}
algorithm to release the excessive load at the sending server
before they reach the receiving server; the third is minimization of the
essential TCP flow control buffer - the socket receive buffer, 
to both enable timely feedback and avoid long queueing delay.
Implementation of our mechanisms is extremely simple without requiring
any kernel or protocol level modification. Our mechanisms work best
for the SIP overload scenarios commonly seen in core networks, where a
small to moderate number of SEs may simultaneously
overload an RE. 
For other scenarios where a large number of SEs overload the RE,
deploying our mechanism will still improve 
performance, but the degree of effectiveness is inherently constrained
by the per-connection TCP flow control mechanism itself. Since each SE
adds to the number of connections and subsequently to the total size
of allocated connection buffers at the RE, as the buffer size
accumulates, so does the delay. Indeed, the solution to this
numerous-SE-single-RE overload problem may ultimately require a shift
from the current push-based model to a poll-based model. Specifically,
instead of allowing all the SEs to send, the RE may advertise a zero
TCP window to most of the SEs and open the windows only for those SEs
that the RE is currently polling to accept loads. Future work is
needed in this area.

Our study sheds light both at software level and conceptual level. At the
software level, we discover implementation flaws for overload
management that would not be noticed without conducting a systematic
overload study, even though our evaluated SIP server is a mature open source
server. At the conceptual level, our results suggest an augmentation to the
long-held notion of TCP flow control: the traditional TCP flow-control
alone is incapable of handling SIP-like time-sensitive session-based application
overload. The conclusion may be generalized to a much broader
application space that share similar load characteristics,
such as database systems. Our proposed combined techniques including {\em connection
split}, {\em smart forwarding} and {\em buffer minimization} are key elements to
make TCP flow control actually work for managing overload of such applications.

\section{Acknowledgement} \label{sec:ack}

The authors would like to acknowledge NTT for funding this project and
Dr. Arata Koike for useful discussions. We would also like to thank
the anonymous reviewers for the helpful comments.

{
\small
\bibliographystyle{plain}
\bibliography{paper}
}

\end{document}